\documentclass[useAMS,usenatbib,a4paper]{mn2e}

\usepackage{epsfig}

\title[Thermal Dust in $z\sim5$ LBGs]{Constraining the Thermal Dust Content of Lyman-Break Galaxies in an Overdense Field at $\mathbf{z\approx5}$}
\author[E R Stanway et al.]{Elizabeth R.~Stanway$^{1}$\thanks{email: E.R.Stanway@Bristol.ac.uk}\thanks{Based on observations collected at the European Organisation for Astronomical Research in the Southern Hemisphere, Chile, associated with observing programmes 175.A-0706 and 083.A-0199.}, Malcolm N.~Bremer$^{1}$, Luke J.~M.~Davies$^{1}$, Matthew D.~Lehnert$^2$
\\$^1$H H Wills Physics Laboratory, Tyndall Avenue, Bristol, BS8 1TL, UK
\\$^2$Laboratoire d'Etudes des Galaxies, Etoiles, Physique et Instrumentation GEPI, Observatoire de Paris, UMR8111 du CNRS, Meudon, 92195 France}

\begin{document}

\date{Accepted  Received ; in original form }

\pagerange{\pageref{firstpage}--\pageref{lastpage}} \pubyear{2010}

\maketitle

\label{firstpage}

\begin{abstract}
We have carried out 870\,$\mu$m observations in the J1040.7-1155
field, known to host an overdensity of Lyman break galaxies at
$z=5.16\pm0.05$.  We do not detect any individual source at the
$S_{870\mu\mathrm{m}}=3.0$\,mJy\,/\,beam (2\,$\sigma$) level. A stack of nine
spectroscopically confirmed $z>5$ galaxies also yields a
non-detection, constraining the submillimeter flux from a typical
galaxy at this redshift to $S_{870\mu\mathrm{m}}<0.85$\,mJy, which
corresponds to a mass limit
M$_\mathrm{dust}<1.2\times10^{8}$\,M$_\odot$ (2\,$\sigma$). This
limits the mass of thermal dust in distant Lyman break galaxies to
less than one tenth of their typical stellar mass. We see no evidence
for strong submillimeter galaxies associated with the
ultraviolet-selected galaxy overdensity, but cannot rule out the
presence of fainter, less massive sources.

\end{abstract}

\begin{keywords}
galaxies: high-redshift -- galaxies: starburst -- submillimetre: galaxies
\end{keywords}

\section{Introduction}
\label{sec:intro}

Thermal dust, interstellar gas and stellar mass all contribute to the
baryon budget of the universe. In the local universe, all three
components can be observed in great detail and the interplay between
them studied.  However, at high redshifts ($z>5$) only the most
luminous, rest-frame ultraviolet-selected stellar component has been
subject to detailed scrutiny.

Observations of the dust content of $z\ga5$ galaxies have been
dominated by studies of quasar hosts
\citep[e.g.][]{2008AJ....135.1201W} and submillimeter galaxies
\citep[e.g.][]{2010arXiv1004.4001C} - relatively massive systems for
their redshift and atypical of the star-forming galaxy
population. Individual studies of far-infrared dust emission in more
typical ultraviolet-luminous star-forming galaxies have been limited
to unusually luminous and lensed examples
\citep[e.g.][]{2001A&A...372L..37B}, while photometrically-selected
samples have been used to derive ensemble constraints from stacking
analyses of UV-continuum selected samples at $z\sim3$
\citep{2003ApJ...582....6W} and galaxies selected for Lyman-$\alpha$
emission at $z\sim5.7$ \citep{2007ApJS..172..518C}.

Constraining the dust properties in these rest-ultraviolet selected
populations as a function of redshift is key to understanding the
early evolution of galaxies. The ultraviolet continuum is
highly sensitive to the effects of dust extinction, leading to
uncertainty in derived physical properties. In particular, the
clustering and star formation history of the universe at early times
may be dramatically underestimated if any significant fraction of the
star-forming galaxy population is omitted from rest-ultraviolet
selected samples.

In our ESO Remote Galaxy Survey \citep[ERGS,][]{2009MNRAS.400..561D}, we
have identified and studied the $z\approx5$ `Lyman break galaxy' (LBG)
population, selected for strong rest-frame ultraviolet continuum
emission, in ten widely-separated fields. In two of these,
J1040.7-1155 and J1054.7-1245, we have detected evidence for large
scale structure in the distant universe.  This comprises overdensities
both in the number counts of a photometrically-selected sample and in
the redshift distribution of spectroscopically confirmed galaxies,
which shows a narrow ($\Delta z=0.1$) spike in source counts (Douglas et
al, 2010, submitted).

We have carried out 870\,$\mu$m observations in the J1040.7-1155 field
(which hosts a galaxy overdensity at $z=5.16\pm0.05$)
using the Large Apex BOlometer CAmera (LABOCA) at the Atacama
Pathfinder EXperiment (APEX) telescope at Chajnantor in Chile in
order to both probe the dust mass of the Lyman-break galaxy population
and to test for the presence of ultraviolet-dark but submillimeter luminous
galaxies that may form part of the same large scale structure. The
unusual density of LBGs, and our extensive optical/near-infrared
imaging and spectroscopy in this field, allows a simultaneous
measurement of the cool dust content of a large number of
spectroscopically-confirmed $z=5$ galaxies.

In this paper we present a constraint on the dust mass of $z=5$ 
LBGs. 
Throughout, we adopt a $\Lambda$CDM
cosmology with ($\Omega_{\Lambda}$, $\Omega_{M}$, $h$)=(0.7, 0.3,
0.7).

\section{Observations}
\label{sec:obs}

Observations were carried out using the Large APEX BOlometer CAmera at
APEX, associated with ESO programme 083.A-0199 (PI Stanway). A total
of 11.7\,hrs of data were collected in service mode over the period
2009 May 10th -- 2009 July 29th, of which 9\,hrs were spent on the
target field. Data were taken using the standard spiral mapping pattern on a four-point
raster to increase the area surveyed, and skydip
observations were used to correct for atmospheric
opacity. Precipitable Water Vapour (PWV) measurements at the site were
typically 0.4-0.7\,mm during observations. Pointing was checked once an
hour.  Standard sources CW-LEO, B13134 and
N2071IR were used for pointing and flux calibration. Flux 
uncertainty was typically 3\%.

\begin{figure*}
\includegraphics[width=0.88\columnwidth]{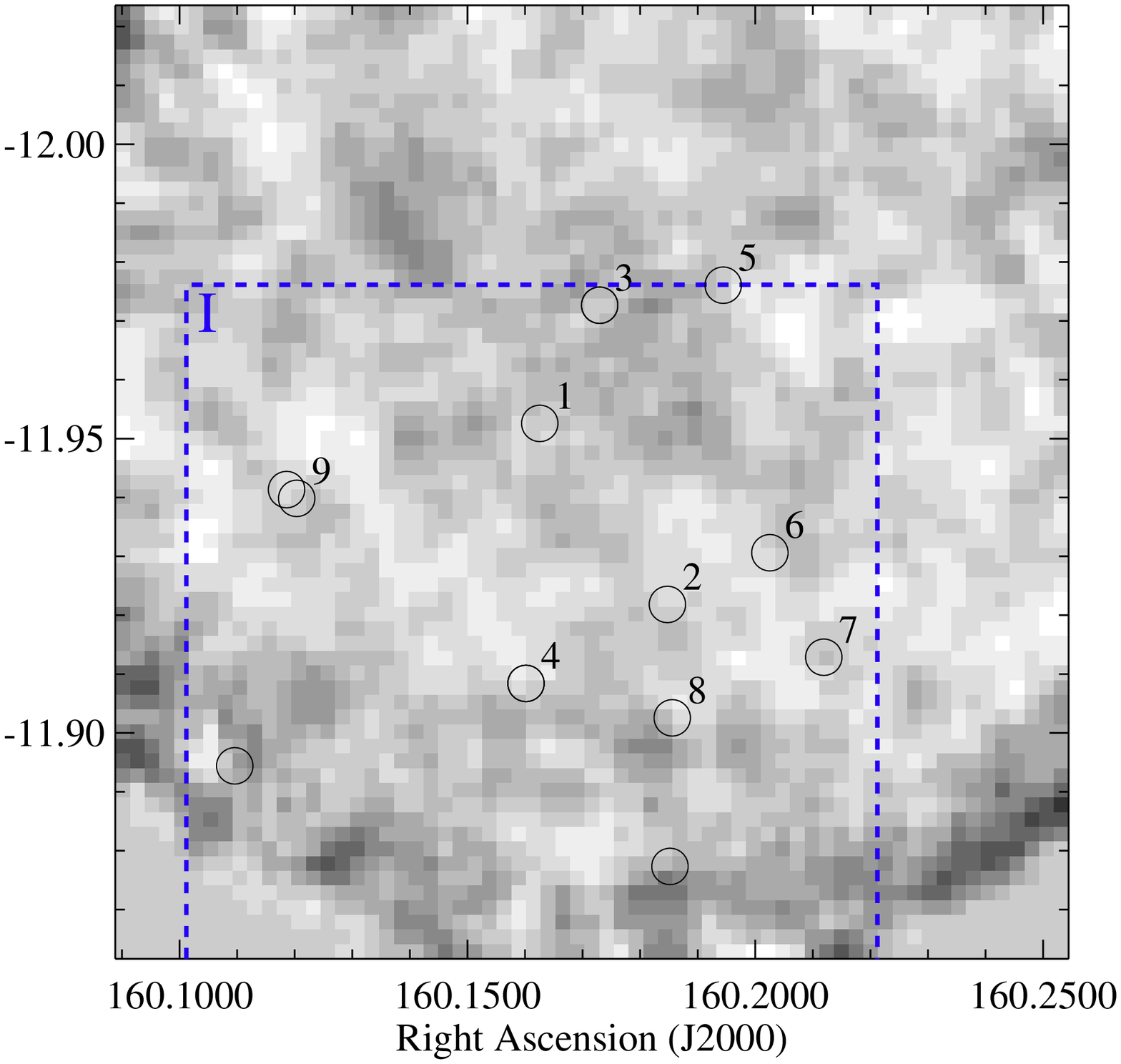}
\hspace{1cm}
\includegraphics[width=0.88\columnwidth]{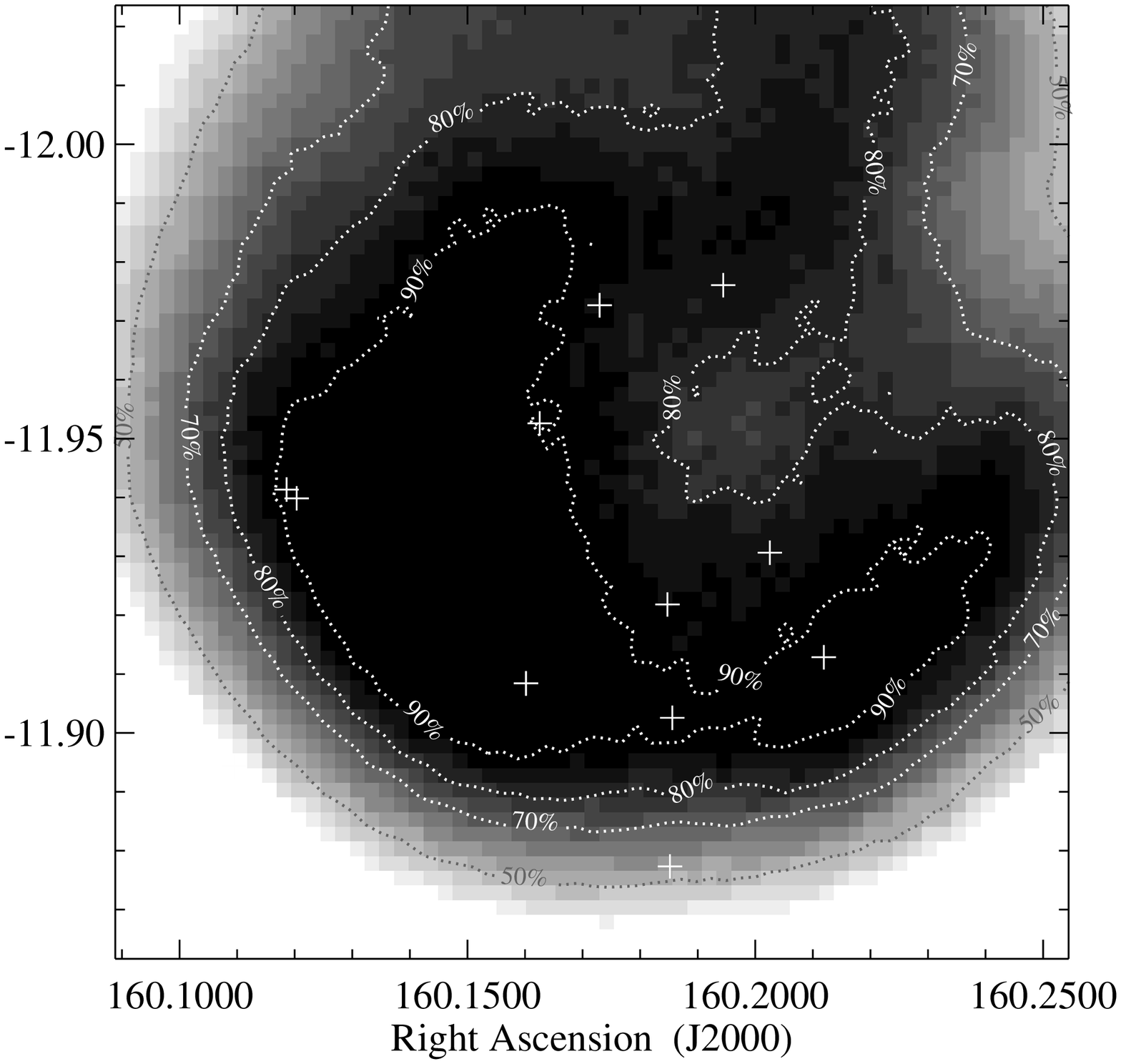}
\caption{Our LABOCA map (left, smoothed by convolution with the 19\arcsec\ beam) and
associated coverage map(right). The dashed line denotes the region
within which we have existing optical data and circles mark the
optical locations of spectroscopically-confirmed $z\sim5$ Lyman break
galaxies. Each circle matches the beam FWHM for an unresolved
source. Numbered circles indicate sources used in the stacking
analysis. Colour is inverted so black indicates a high-flux signal.
\label{fig:images}}
\end{figure*}

The field was centered at
10$^\mathrm{h}$40$^\mathrm{m}$42\fs2\,-11\degr58\arcmin00\arcsec\ (J2000), on a
$z=5.12$ galaxy identified from its CO(2-1) emission in
the rest-frame far-infrared \citep{2008ApJ...687L...1S}. Discussion of
this non-LBG source is deferred to a later paper. Unfortunately the
project was not completed before the end of the semester, leading to
an irregular coverage of the field, with regions of the survey area
(which partly overlaps our optical imaging) significantly deeper
than others.  A total area of approximately $9\farcm7\times9\farcm7$
was surveyed to 50\%, and $7\farcm7\times7\farcm7$ to 75\%, of the
best total coverage (figure \ref{fig:images}).

Data were reduced using the dedicated Bolometer Array (BoA) analysis
software developed to reduce LABOCA data, using an algorithm optimised
to recover weak sources.  LABOCA has a beam size of 18\arcsec\ on the
APEX telescope and the data were resampled to a pixel scale of
9\arcsec\ per pixel during the reduction procedure. In common with
other authors \citep[e.g.][]{2009ApJ...707.1201W}, we
smooth our data by convolution with the beam width in order to optimise
signal to noise for point sources. The root mean square noise on
the smoothed image was
1.49\,mJy/beam (measured over a region with mean coverage 84\%).

\section{A Limit on the Dust Masses}
\label{sec:lbg_dust}

Twelve spectroscopically-confirmed $z>5$ Lyman break galaxies lie
within the field of our LABOCA observations (shown by circles in
figure \ref{fig:images}). Of these, two lie southwards of the field
centre in a region with poor flux sensitivity and are not considered further in
this analysis. Two other sources are separated by a projected angular
distance of only a quarter beam although there is actually a
substantial line of sight distance between them, with one source lying
at $z=5.15$ and the second being a background source at $z=6.41$
(originally selected as an $I$-band dropout galaxy). This is the sole
outlier in our spectroscopically confirmed sample, with the remaining
galaxies lying in the redshift range $5.1<z<5.5$, with a mean redshift
$<z>=5.177$.

None of the confirmed $z=5$ galaxies in this field are individually
detected. Indeed, there is, at most, one convincing 3-5\,$\sigma$ detection
across our field \citep[lying well north of the region surveyed in our
optical imaging and consistent with the 1-2 sources expected based
on other 850\,$\mu$m
surveys,][]{2006MNRAS.372.1621C,2009ApJ...707.1201W}. The 18\arcsec\
beam of LABOCA corresponds to an angular scale of 112\,kpc at
$z=5.18$, and the stellar light of Lyman break galaxies in our sample
is typically extended over a half-light radius of just 0.14\arcsec\
\citep{2009MNRAS.400..561D}. We thus expect any dust emission from our
distant sources to be unresolved in LABOCA imaging, and can place a
2\,$\sigma$ upper limit $S_{870\mu\mathrm{m}}<3.0$\,mJy on the flux
from each galaxy.

We also consider a mean stacked image constructed from a $40\times40$
pixel ($20\times20$ beamsize) region surrounding each of the nine
$z\sim5$ LBGs labelled in figure \ref{fig:images}. Eight of these show
Lyman-$\alpha$ emission lines, although only four have line equivalent
widths $>30$\AA, sufficiently high for detection as Lyman-alpha
emitters (LAEs) in typical narrowband surveys. Since our nine sources
lie in a region with near-constant coverage we use a simple mean;
using a noise-weighted average changes the results by $<$2\%, well
within the statistical errors.  An inspection of this stack also
failed to yield a detection, with an improved upper limit on the
average source of $S_{870\mu\mathrm{m}}<0.85$\,mJy (2\,$\sigma$, based
on the image noise, which is well fit by a Gaussian distribution). We
note that contributions from sources along the line of sight between
us and the $z=5$ target galaxies or from the $z=6.4$ background source
can only increase the measured flux in the stack; hence this is a
strict upper limit. Emission from line-of-sight sources offset from
the targets will increase the noise levels in the image stack and such
confusion may be a limiting factor on our point-source
sensitivity. Again this effect acts to increase rather than decrease
our upper limit.

 Following \citet[][eqns 3 \& 4]{2008A&A...491..173A}, we convert our flux limit into
 a constraint on the dust mass of each galaxy by modelling the
 spectrum as a grey body, taking into account dust heating by the CMB
 as a function of redshift and observing frequency (a significant
 effect at high redshifts). We do not assume an optically-thin approximation 
but use the full expression. In the absence of quantitative information
 about the properties of dust grains at very high redshift we use the
 same far-infrared dust emissivity coefficient used by
 \citet{2008A&A...491..173A}.  The power law emissivity index is fixed
 to $\beta=2$, found to fit a sample of $z>4$ quasars by
 \citet{2001MNRAS.324L..17P} and rather steeper than the $\beta=1.6$
 found for lower redshift quasars by \citet{2006ApJ...642..694B}.
 This steeper emissivity law is appropriate for relatively small dust
 grains \citep{1990A&A...237..215D} and hence may arise from evolution
 in the dust properties of galaxies at high
 redshift. \citet{2004Natur.431..533M} found that rest-frame
 ultraviolet dust absorption in $z>6$ sources is best
 described by smaller grains with production dominated by Type-II SNe,
 an effect also seen in $z>5$ gamma ray burst host galaxies
 \citep{2007ApJ...661L...9S,2009arXiv0912.2999P}.  Altering the
 emissivity index from $\beta=2$ to $\beta=1.6$ increases the derived
 dust masses by a factor of 2.

\begin{figure}
\includegraphics[width=0.98\columnwidth]{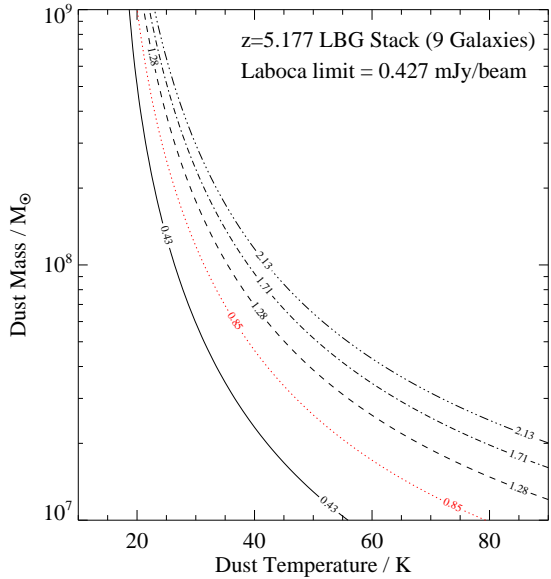}
 \caption{Constraints on the dust mass derived from our 870\,$\mu$m
 flux limit for a stack of nine $z=5$ Lyman break galaxies. Contours
 show dust masses excluded at 1, 2, 3, 4 and 5\,$\sigma$ level as a
 function of black body temperature.}
 \label{fig:dustlims}
 \end{figure}

 Figure \ref{fig:dustlims} shows our constraint on the dust mass of a
 typical galaxy in our sample as a function of assumed dust
 temperature. The LABOCA non-detection limit from our LBG stack either
 constrains the dust temperature in a typical UV-luminous galaxy at
 $z=5.18$ to $<30$\,K or the dust mass to
 M$_\mathrm{dust}<1.2\times10^{8}$\,M$_\odot$ (2\,$\sigma$), with
 constraints on individual galaxies being a factor of three times
 weaker.

 The highly lensed $z=2.7$ Lyman break galaxy MS1512-cB58 has a dust
 temperature T$=33$\,K based on fitting of its submillimeter spectral
 energy distribution \citep{2001A&A...372L..37B}.  The mean dust
 temperature of submillimeter selected galaxies at $z\sim2$ is
 comparable: $\sim35$\,K
 \citep{2005ApJ...622..772C,2006ApJ...650..592K}. Our galaxies are
 less massive, and their integrated star formation rates lower, than
 typical submillimeter galaxies. However, the star formation intensity
 in their central regions is comparable -- a few tens of solar masses
 per year per square kiloparsec \citep{verma07}.  Hence we adopt 30\,K as an estimate
 for dust temperature (and the resulting mass constraint) in the
 discussion that follows.

At this temperature, assuming the emissivity index and dust model
discussed above,
$L_\mathrm{FIR}=2.8\times10^{12}$\,$S_{870\mu\mathrm{m}}$ for a source
at $z=5.16$, where $L_\mathrm{FIR}$ is the integrated far-infrared
luminosity integrated between 40 and 120\,$\mu$m in the rest-frame and
measured in solar luminosities \citet{1992ARA&A..30..575C} and
$S_{870\mu\mathrm{m}}$ is the observed flux at 870\,$\mu$m in
Janskys. Hence our flux limit implies a typical $z\sim5$ LBGs has
$L_\mathrm{FIR}<2.4\times10^9\,L_\odot$. We do not extrapolate further
to obtain a star formation rate from this value, since the assumptions
required to do so at $z\sim5$ are far from clear. As \citet{carilli08}
discuss, the far-infrared to radio luminosity correlation is expected
to break down at high redshifts, where inverse Compton scattering off
the cosmic microwave background is likely to be a significant effect.

\section{Discussion}
\label{sec:discussion}

\subsection{LBGs at $z\sim5$}
\label{sec:lbgs-at-z5}

This work provides the first constraint on thermal dust emission from
spectroscopically-confirmed, rest-ultraviolet continuum selected
galaxies at $z\sim5$. However, \citet{2007ApJS..172..518C} derived an
upper limit on the 1.2\,mm flux of a sample of ten galaxies selected
to have a flux excess consistent with strong Lyman-$\alpha$ line emission at
$z\sim5.7$. Applying the same analysis used here and assuming
T$=30$\,K, their stacking analysis limit on the typical flux of these
sources, $S_\mathrm{1.2mm}<0.7$\,mJy (2\,$\sigma$), corresponds to a
dust mass limit M$_\mathrm{dust}<1.9\times10^{8}$\,M$_\odot$.
Galaxies selected as LBGs are typically more luminous in the
rest-frame ultraviolet continuum and hence believed to be older and
more massive than similar sources selected for a very high Lyman-$\alpha$
emission line equivalent width \citep[e.g.][]{2008ApJ...674...70L}. 
However, the dust-mass constraint in
this paper is now within 30\% of the tight limit on dust mass in
$z=5.7$ Lyman-$\alpha$ emitters, suggesting that neither of these
populations hosts any appreciable dust component.

The stellar masses of Lyman break galaxies at $z\sim5$ are typically
$\sim$ few $\times10^{9}$\,M$_{\odot}$
\citep{verma07,2009ApJ...697.1493S}. Our limit on the dust mass of
these galaxies M$_\mathrm{dust}<1.2\times10^{8}$\,M$_\odot$ at 30\,K
constrains their baryonic mass component in the form of thermal dust to
be no more than about a tenth of the stellar component.

Interestingly, this upper limit is already comparable to the stellar
mass to warm dust ratio observed in starbursts in the local universe.
\citet{2005A&A...434..849H} studied dust emission from a small sample
nearby blue compact dwarfs, which are comparable in both metallicity
and star formation intensity to $z\sim5$ LBGs, albeit less massive,
and have emission dominated by small-grained, supernova-formed dust
much like galaxies at $z>5$. From SED fitting of the starbursts,
\citeauthor{2005A&A...434..849H} derived a stellar mass to dust mass
ratio of 10 and 13 for two sources with Z/Z$_\odot$=0.14 and 0.20
respectively. The metallicity of $z\sim5$ LBGs is still poorly
constrained, although estimates in the range 0.1-0.2\,Z$_\odot$ appear
appropriate from SED fitting \citep[e.g.][]{verma07} and analysis of
the rest-UV spectral slope (Douglas et al, 2010).

Lyman break galaxies at $z\sim3$ are selected for rest-frame
ultraviolet continuum in the same manner as the $z\sim5$ LBG sample
discussed here, and hence have comparable star formation rates, and yet
differ significantly in their physical properties. Lyman break
galaxies at $z\sim3$ are typically older and more massive than those
at $z\sim5$ (having had longer to evolve). They have similar or
slightly higher metallicities and also appear to have slightly larger
dust extinction in the rest-frame ultraviolet \citep{verma07}. Despite
these differences, the well-studied $z\sim3$ population provides the
most direct comparison set with UV-continuum selected samples at
higher redshift.

Individual detections of thermal dust emission have now been made in
several $z\sim3$ star-forming galaxies, generally made possible by a
high degree of gravitational lensing. The well-known $z=2.7$ source
MS1512-cB58 hosts a dust mass of $1.2\times10^7$\,M$_\odot$
\citep[based on the 1.2\,mm flux of][ using T=30\,K, $\beta=2$ and
correcting for lensing $\mu=32$]{2001A&A...372L..37B} and a stellar
mass of $\sim10^{9}$\,M$_\odot$ \citep{2008ApJ...689...59S}. Similarly
the $z=3.07$ Cosmic Eye, hosts a stellar mass
M$^\ast=6\times10^9$\,M$_\odot$ and an estimated dust mass
M$_\mathrm{dust}=2.5\times10^{7}$\,M$_\odot$
\citep{2007ApJ...665..936C}. These very low dust masses are consistent
with an ensemble average estimated from a stack of submillimeter
fluxes for photometrically-selected $z\sim3$ Lyman break galaxies from
the Canada-UK Deep Submillimeter Survey. \citet{2003ApJ...582....6W}
measured a 2\,$\sigma$ upper limit on the flux of a typical $z\sim3$
LBG, $S_{850\mu\mathrm{m}}<0.4$\,mJy, corresponding to a dust mass
M$_\mathrm{dust}<5\times10^{7}$\,M$_\odot$ at 30\,K using the
cosmology in this paper.

If the relatively low dust masses seen in $z\sim3$ LBGs are common at
higher redshifts, then significantly deeper submillimeter data will be
required to detect the dust emission from $z\sim5$ sources. However
our limit is already sufficient to challenge any paradigm in which
a significant amount of the star formation at $z\sim5$ is heavily
obscured. 

This argues against models in which the multi-component morphology of
many 
rest-ultraviolet selected high reshift galaxies is 
analogous to the distribution of
massive starburst regions seen in local
ultraviolet-luminous galaxies, while the bulk of the underlying galaxy
remains unseen due to heavy obscuration. The high far-infrared fluxes
predicted for such embedded obscured clumps are hard to reconcile with
the tight submillimeter constraints emerging for high redshift
populations \citep{2008ApJ...677...37O,2009ApJ...706..203O}.

\subsection{Structure in the J1040.7-1155 Field}
\label{sec:1040-1155}

While the photometric properties of individual Lyman break galaxies
contributing to our analysis are rather typical of $z\sim5$ Lyman
break galaxies as a whole (Douglas et al, 2010), it is worth noting
that the large scale structure properties of the field are far from
typical. The redshift distribution of spectroscopically confirmed
galaxies in the J1040.7-1155 field shows a 6\,$\sigma$ peak relative
to the mean distribution in our spectroscopic survey, lying at
$z=5.16\pm0.05$. The field is also overdense in photometrically
selected $z\sim5$ candidates and hosts an ultraviolet-dark $z=5.12$
galaxy with a large mass of gas in the form of carbon monoxide
\citep[and, by inference, molecular hydrogen,][ 2010 in
prep]{2008ApJ...687L...1S}.  Six of the galaxies
contributing to this analysis form part of the large
scale structure in this field.

Two of these sources, objects 3 and 5 in figure \ref{fig:images}, have
tight constraints on their molecular gas masses:
M(H$_2$)$<1.7\times10^{10}$ and M(H$_2$)$<2.9\times10^{9}$\,M$_\odot$
(2\,$\sigma$) respectively, based on measurements of carbon monoxide
emission at millimeter wavelengths \citep{2008ApJ...687L...1S}.  Object 3 is confused with a bright
neighbouring source in our deep {\em Spitzer}/IRAC imaging of
this field, but otherwise typical of Lyman
break galaxies in our sample. In the case of
object 5 we are also able to constrain its rest-frame optical
flux based on non-detection in our IRAC imaging and estimate 
its stellar mass M$^\ast\sim2.5\times10^{9}$
(Stanway et al, in prep).   
In this source, a confirmed $z=5.116$ galaxy, the stellar
mass contributes a minimum of half of the observed baryonic content of
the galaxy, with the mass in molecular gas potentially comparable to
this, and a contribution from thermal dust $<10$\% of the observed
total.

These data allow us not only to study the properties of known
rest-frame ultraviolet luminous sources, but also to determine whether
massive, dusty sources exist within the same large scale structure at
very early times. 
The excess of Lyman-break selected galaxies may well be tracing out
massive large scale structures in which the majority of baryonic
material is undetected in the ultraviolet \citep{2008ApJ...687L...1S}.
If massive submillimeter galaxies preferentially occupy strong peaks
in the matter density distribution \citep{blain04}, we might expect to
detect an excess such galaxies in this field. Such behaviour has been
observed using LABOCA by \citet{beelen08} who identified a possible
overdensity of submillimeter galaxies apparently associated with
strong Lyman-$\alpha$ emitting sources forming a protocluster at
$z=2.4$, by \citet{2000ApJ...542...27I} who observed massive
SCUBA galaxies clustering with Lyman break galaxies in the
environs of a $z=3.8$ radio galaxy, and by \citet{Aravena10} who observed
BzK galaxies clustering with MAMBO-selected sources at $z\sim2$. Given the current lack of
information on typical luminous lifetimes, dust properties, clustering
scales and mass scales of $z\sim5$ LBGs, submillimeter and
gas-dominated galaxies, it is impossible to make a quantitative
prediction for the number of submillimeter sources expected in an
overdensity of Lyman break galaxies. However there is no evidence
for this field being overdense in submillimeter galaxies at any redshift.

At dust temperatures of $\sim$30-35\,K the lack of 3\,$\sigma$
detections for continuum sources in our field implies a dust mass
limit of $\sim2-4\times10^8$\,M$_\odot$ at $z\sim5$, ruling out the
presence of a massive UV-obscured active galaxy or submillimeter
galaxy comparable to others seen at high redshift \citep[see
compilation in][]{2005ARA&A..43..677S}. However, we caution that our
constraints on individual sources are weak (and depend on the assumed
dust temperature).  We cannot rule out a substantial population of
cooler, less massive or less dusty sources in this field. The apparent
lack of massive evolved systems at this redshift may indicate that
galaxies, even those in apparently overdense regions, may have a more
complex history than simply growing into massive galaxies observed
locally.

\section{Conclusions}
\label{sec:conc}

We describe 870\,$\mu$m observations undertaken using LABOCA  and targetting a field known to host an overdensity of
Lyman break galaxies at high redshift. Our main conclusions can be summarised as follows:

\begin{enumerate}

\item No individual galaxy is detected in our LABOCA observations, to a
a 2\,$\sigma$ upper limit $S_{870\mu\mathrm{m}}<3.0$\,mJy. 

\item By stacking nine spectroscopically confirmed $z\sim5.2$ galaxies
in this field, we are able to constrain the thermal dust emission in a
typical Lyman break galaxy at this redshift to
$S_{870\mu\mathrm{m}}<0.85$\,mJy, corresponding to a dust mass
 M$_\mathrm{dust}<1.2\times10^{8}$\,M$_\odot$ (2\,$\sigma$) at a dust
temperature of 30\,K.

\item These observations place an upper limit on the thermal dust fraction
 in a typical LBG at $z\sim5$ of $\sim10$\% of the stellar mass.

 \item We see no evidence for an excess of heavily-obscured massive
starburst galaxies associated with a large scale structure at $z=5.16$
in this field.

\end{enumerate}

Deeper submillimeter observations are required to
better constrain the dust properties of galaxies at high redshift. Our
observational programme to characterise $z\sim5$ galaxies is
continuing, with particular emphasis on both this field and a second
overdense region. We note that the Atacama Large Millimeter Array
(ALMA) will have a transformative effect on these studies,
making dust continuum detections in typical, unlensed Lyman
break galaxies accessible for the first time. However  the
small field of view of ALMA at submillimeter wavelengths (just
13\,arcseconds at 250\,GHz) will reduce its effectiveness for large
scale mapping of the relatively sparse high redshift galaxy
population.

\section*{Acknowledgments}
ERS and LJMD gratefully acknowledge support from the UK Science and
Technology Facilities council. This publication is based on data
acquired with the Atacama Pathfinder Experiment (APEX). APEX is a
collaboration between the Max-Planck-Institut fur Radioastronomie, the
European Southern Observatory, and the Onsala Space Observatory.

\bsp

\label{lastpage}

\end{document}